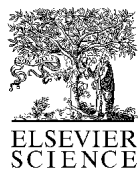
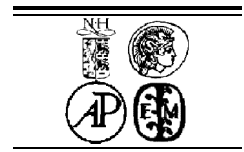

# Persistent magnetoresistive memory in phase separated manganites

P. Levy,[a,*] F. Parisi,[a,#] J. Sacanell,[a] L. Granja,[a] M. Quintero,[a] J. Curiale,[a] G. Polla,[a] G. Leyva,[a] R.S. Freitas,[b] L. Ghivelder,[b] C. Acha,[c] T. Y. Koo,[d] and S. -W. Cheong[d]

[a] *U.A.Física, CAC, CNEA, San Martín, Pcia. Buenos Aires, Argentina*

[b] *Instituto de Física, Universidade Federal do Rio de Janeiro, Rio de Janeiro, Brazil*

[c] *Lab. Bajas Temperaturas, FCEyN, UBA, Buenos Aires, Argentina*

[d] *Department of Physics and Astronomy, Rutgers University, New Jersey, USA*

**Abstract**

We have studied magnetic and transport properties on different manganese - oxide - based compounds exhibiting phase separation: polycrystalline $La_{5/8-y}Pr_yCa_{3/8}MnO_3$ (y=0.3) and $La_{1/2}Ca_{1/2}Mn_{1-z}Fe_zO_3$ (z = 0.05), and single crystals of $La_{5/8-y}Pr_yCa_{3/8}MnO_3$ (y~0.35). Time dependent effects indicate that the fractions of the coexisting phases are dynamically changing in a definite temperature range. We found that in this range the ferromagnetic fraction "*f*" can be easily tuned by application of low magnetic fields (< 1 T). The effect is persistent after the field is turned off, thus the field remains imprinted in the actual value of "*f*" and can be recovered through transport measurements. This effect is due both to the existence of a true phase separated equilibrium state with definite equilibrium fraction "$f_0$", and to the slow growth dynamics. The fact that the same global features were found on different compounds and in polycrystalline and single crystalline samples, suggests that the effect is a general feature of some phase separated media. © 2001 Elsevier Science. All rights reserved

The simultaneous presence of submicrometer ferromagnetic metallic (FM) and charge ordered and/or paramagnetic insulating regions observed in some manganese-oxide-based compounds, the phase separation (PS) phenomena,[1,2] is a major discovery in the study of strongly correlated electron systems.[3,4] There are conclusive evidences, both from theoretical[1,3,4] and experimental[5-7] works showing that the high values of low field magnetoresistance achieved in PS manganites are due to the unique possibility of unbalancing the amount of the coexisting phases by the application of low magnetic fields.

In PS compounds with their FM ordering temperature $T_C$ lower than the charge/orbital order transition temperature ($T_{co}$), the state close below $T_C$ is characterized by the coexistence of isolated FM clusters embedded in insulating regions. These clusters grow as the temperature is lowered, and the insulating to metal transition is obtained when the fraction of the FM phase reaches the percolation threshold.[2] Time dependent effects related to the slow growth of the FM phase in PS media are well documented in films [8], single [9-11] and polycrystalline [12] samples.

---

[*] Corresponding author. Tel.(+54-11)-6772-7104 ; fax: (+54-11)-6772-7121; e-mail: levy@cnea.gov.ar (P.Levy)

We have recently shown that the actual amount of the FM phase can be tuned by means of a low magnetic field[13] in polycrystalline $La_{0.325}Pr_{0.300}Ca_{0.375}MnO_3$, which acts as a nonvolatile memory. In this work we extend those findings to other compounds belonging to the "low-$T_C$" family, i.e. compounds with $T_C < T_{co}$. The fact that the same global features were found on different compounds, and in polycrystalline and single crystalline samples, suggest that the effect is a distinctive feature of low-$T_C$ phase separated media.

We studied the time dependent response on different compounds exhibiting the phase separation phenomena: polycrystalline $La_{5/8-y}Pr_yCa_{3/8}MnO_3$ (y=0.3, grain size 2 µm) and $La_{1/2}Ca_{1/2}Mn_{1-z}Fe_zO_3$ (z= 0.05, grain size ~ 0.6 µm), both of them synthesized by thermal decomposition of the corresponding citrates followed by thermal treatments, and single crystalline $La_{5/8-y}Pr_yCa_{3/8}MnO_3$ (y~0.35), grown by the travelling floating zone method. We used transport (standard 4 probe technique) and magnetization (extraction QD PPMS magnetometer) measurements to characterize the dynamics of the coexisting phases, which is observed close below $T_C$.

We have measured the time evolution of the isothermal resistivity $\rho(t)$ of each sample at selected temperatures, corresponding to the region in which the systems display out of equilibrium features, characterized by the isothermal increase of the size of the FM regions. A nearly logarithmic decrease of $\rho$ and a concomitant increase of the magnetization (measured with a low field, i.e. 0.001 T) was observed.

Figures 1, 2 and 3 display the normalized resistivity $\rho(t)/\rho(0)$ measured as a function of time for polycrystalline $La_{0.5}Ca_{0.5}Mn_{0.95}Fe_{0.05}O_3$, and $La_{0.325}Pr_{0.300}Ca_{0.375}MnO_3$, and single crystalline $La_{0.275}Pr_{0.350}Ca_{0.375}MnO_3$ respectively. To obtain the data we first cooled the samples in zero field down to the target temperature, and then the isothermal evolution of the resistivity was acquired, while an external low magnetic field H < 1 T was applied and removed. Some common features to all the samples emerge, namely:

- jumps are observed in $\rho(t)$ when the field is applied and removed, related to the alignment of spins and domains and to the enlargement of the FM phase induced by the applied H;
- a time dependence of $\rho(t)$ is observed while H remains applied, attributed to the slow isothermal enlargement of the FM phase reinforced by the external stimulus provided by the field;
- the remnant resistivity ($\rho_{H=0}(t)$, i.e. $\rho(t)$ when H=0) is determined by the previously applied H (the dependence is on both the intensity and the accumulated time).
- time relaxation of the remnant resistivity was observed, although this fact is not apparent in all graphs due to time scales used. The possibility to determine the sign of this relaxation with the magnitude of H (i.e. increase or decrease of $\rho_{H=0}(t)$ after applying and removing certain H) signs the existence of a true PS state characterized by an equilibrium fraction "$f_0$" of the FM phase.[13,14]

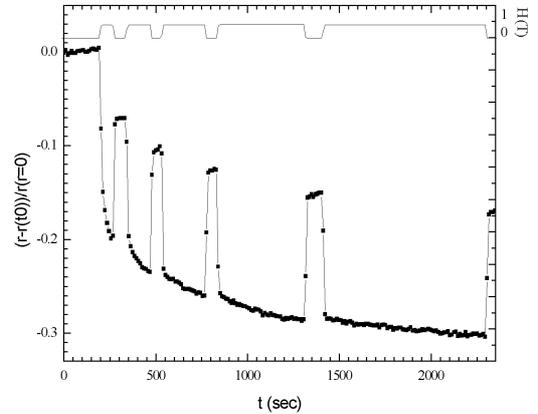

Fig. 1. Normalized resistivity as a function of time after zero field cooling down to T= 81.1 K, for polycrystalline $La_{0.5}Ca_{0.5}Mn_{0.95}Fe_{0.05}O_3$. The applied field value is H= 0.4 T during time intervals of 1, 2, 4, 8 and 16 minutes.

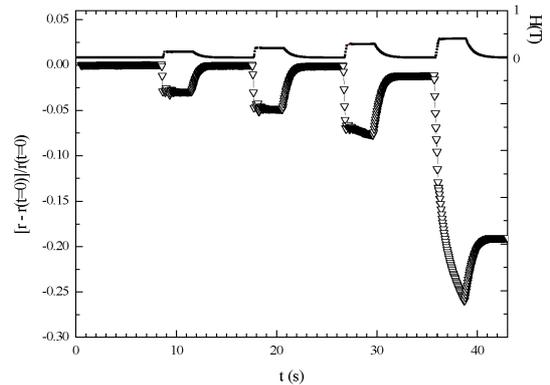

Fig. 2. Normalized resistivity as a function of time after zero field cooling down to T= 94.5 K, for polycrystalline $La_{0.325}Pr_{0.300}Ca_{0.375}MnO_3$. Applied field values were H= 0.1, 0.2, 0.3 and 0.4 T for 3 seconds, with time intervals of 6 seconds.

The combination of alignment and enlargement effects determines a response which is persistent after the magnetic field is removed, remains encoded in the relative fraction "*f*" of the FM phase, and can be recovered through transport techniques.

Measurements performed on different time scales are shown so as to depict the generality of the observed phenomena. The experiment shown in Fig. 1 allows us to identify the time scale of the growth mechanism of the FM regions, ranging up to several minutes. In Fig. 2 the possibility to determine the relative fractions after applying the external field for only 3 seconds is apparent. We emphasize the fact that a magnetic field as low as 0.2 T is able to modify in a sizeable and persistent way the $\rho_{H=0}$ value of the virgin samples, both in poly and single crystalline materials.

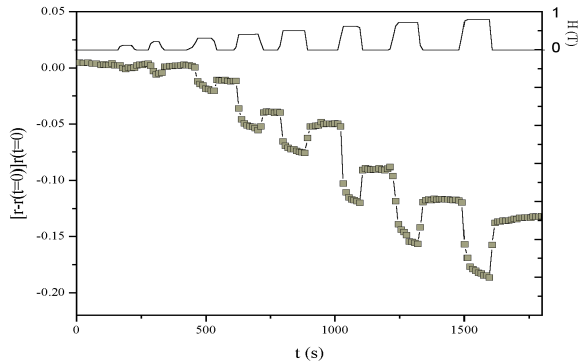

Fig. 3. Normalized resistivity as a function of time after zero field cooling down to T= 91.1 K, for single crystalline $La_{0.275}Pr_{0.350}Ca_{0.375}MnO_3$. Applied magnetic field values were H= 0.1, 0.2, 0.3, 0.4, 0.5, 0.6, 0.7 and 0.8 T.

The memory effect is due both to the existence of a true phase separated equilibrium state with definite equilibrium fractions, and to the capability of the external perturbation to tune different metastable states. This persistence is protected by the slow dynamics displayed by the systems in the temperature range where out of equilibrium features are present. The above described scenario seems to be characteristic of the so called "low $T_C$" manganites exhibiting PS, i.e intrinsically inhomogeneous media with $T_C$ lower than the charge ordering temperature, irrespective of the hole doping level. The fact that the same global features were found in polycrystalline and single crystalline samples suggests that the effect is not produced by extrinsic properties related to grain boundaries, but is a distinctive feature of the studied PS compounds. Recent studies[12,15] suggest that the common feature underlying the PS phenomena are strain states localized at the FM – charge ordered boundaries, which are originated in the martensitic - like nature of the growth of a child phase within the parent one.

Summarizing, the capability of small external forces to tailor the macroscopic response of the material is a unique possibility related to the intrinsic submicrometer coexistence of regions with very different transport and magnetic properties. This fact implies the feasibility to unbalance the amount of coexisting phases with relatively small perturbations, producing relatively huge changes in certain described conditions. As the effect of this changes is persistent and can be recovered through electrical transport techniques, potential applications of these materials can be envisaged.

Project partially financed by Fundación Vitae. Support from Agilent Technologies Argentina is acknowledged.

[*]Member of CONICET, also at U.N. de Quilmes.
[#]Also at E.C.yT., U.N. de San Martín.